\documentclass[twocolumn,aps,pra,showpacs]{revtex4}

\begin{document}
\title{Loophole-free Bell's experiments and two-photon
all-versus-nothing violations of local realism}
\author{Ad\'{a}n Cabello}
\email{adan@us.es}
\affiliation{Departamento de F\'{\i}sica Aplicada II,
Universidad de Sevilla, 41012 Sevilla, Spain}
\date{\today}


\begin{abstract}
We introduce an extended version of a previous all-versus-nothing
proof of impossibility of Einstein-Podolsky-Rosen's local elements
of reality for two photons entangled both in polarization and path
degrees of freedom [A. Cabello, Phys. Rev. Lett. {\bf 95}, 210401
(2005)], which leads to a Bell's inequality where the classical
bound is 8 and the quantum prediction is 16. A simple estimation
of the detection efficiency required to close the detection
loophole using this extended version gives $\eta > 0.69$. This
efficiency is lower than that required for previous proposals.
\end{abstract}


\maketitle


If, from the result of one experiment, we can predict with
certainty the result of a spacelike separated experiment, then,
following Einstein, Podolsky, and Rosen (EPR)~\cite{EPR35}, there
must be a local element of reality (LER) corresponding to the
latter result. However, some predictions of local realistic
theories are in conflict with those of quantum
mechanics~\cite{Bell64,CHSH69}.
Experiments~\cite{FC72,WJSHZ98,RKVSIMW01} have shown an excellent
agreement with quantum mechanics and have provided solid evidence
against LERs. So far, however, the results of these experiments
still admit an interpretation in terms of LERs. A conclusive
loophole-free experiment would require spacelike separation
between the local experiments and a sufficiently large number of
the prepared pairs' detections; otherwise, the possibility of
communication at the speed of light between the particles cannot
be excluded (locality loophole~\cite{Bell81}), and neither can it
be excluded that the nondetections correspond to local
instructions like ``if experiment $X$ is performed, then do not
activate the detector'' (detection loophole~\cite{Pearle70}).

Photons are the best candidates for closing the locality loophole.
For instance, the Innsbruck experiment~\cite{WJSHZ98} with
polarization-entangled photons separated $400$\,m is not subject
to the locality loophole; however, its detection efficiency ($\eta
= 0.05$) was not high enough to close the detection loophole
($\eta \ge 0.83$ is required~\cite{CH74}). The detection
efficiency for ions is much higher. For instance, in the Boulder
experiment~\cite{RKVSIMW01} with trapped beryllium ions, $\eta
\approx 0.98$; however, the distance between ions ($3\,\mu$m) was
not enough to close the locality loophole.

There are several proposals for experiments for closing both
loopholes~\cite{LS81}; however, most of them are very difficult to
implement with current technology. The most promising approach for
a loophole-free experiment is by using entangled photons and more
efficient photodetectors~\cite{KESC94}. Recent experiments with
pairs of entangled photons have achieved $\eta =
0.33$~\cite{Kwiat05}. Closing the detection loophole with
maximally entangled states and the Clauser-Horne-Shimony-Holt
(CHSH) Bell's inequality~\cite{CHSH69} requires a detection
efficiency $\eta > 2 (\sqrt{2}-1) \approx 0.83$~\cite{CH74}. By
using nonmaximally entangled states and supplementary
assumptions, $\eta$ can be lowered to $\eta >
0.67$~\cite{Eberhard93}. However, these experiments are based on a
different interpretation of EPR's condition for LERs~\cite{CG97}.

The detection efficiency required for a loophole-free experiment
on Bell's theorem of impossibility of EPR's LERs is related with
the statistical strength of the proof tested in the experiment
(i.e., with the amount of evidence against LERs provided by the
corresponding experiment). In this respect, all-versus-nothing
(AVN) proofs~\cite{GHZ89,Cabello01a} provide stronger evidence
against LERs than other proofs~\cite{VGG03}. Specifically, a
loophole-free experiment based on the three-observer
version~\cite{Mermin90a} of Greenberger, Horne, and Zeilinger's
proof~\cite{GHZ89}, would require $\eta > 0.75$~\cite{Larsson98}.
The negative side is that it requires three spacelike separated
regions. The two-photon version~\cite{CPZBZ03,YZZYZZCP05,BCDM05}
of the two-observer AVN proof~\cite{Cabello01a} only requires
a spacelike separation between two regions, but the detection
efficiency needed for a loophole-free test is $\eta > 5/6 \approx
0.83$~\cite{Cabello05a}.

In this paper we introduce a new AVN proof for two photons, and
its corresponding Bell's inequality, which requires an efficiency
$\eta > 0.69$ to close the detection loophole. This efficiency,
although still higher than that achieved in recent experiments, is
lower than that required for any previous proposal for a
loophole-free experiment based on bipartite Bell's inequalities
and the usual interpretation of EPR's condition. The new AVN proof
is an extended version of a previous one~\cite{Cabello05a}.

Consider two photons entangled both in polarization and in path
degrees of freedom~\cite{CPZBZ03,YZZYZZCP05,BCDM05,Kwiat97}
prepared in the state
\begin{eqnarray}
|\psi\rangle & = & \frac{1}{2} ( |H u\rangle_1 |H u\rangle_2 +
|H d\rangle_1 |H d\rangle_2 \nonumber \\
& & + |V u\rangle_1 |V u\rangle_2 - |V d\rangle_1 |V d\rangle_2),
\label{benasque}
\end{eqnarray}
where $|H\rangle_j$ and $|V\rangle_j$ represent horizontal and
vertical polarization, and $|u\rangle_j$ and $|d\rangle_j$ denote
two orthonormal path states for photon-$j$. Consider also six
local observables on photon-$j$: three for polarization degrees of
freedom, defined by the operators
\begin{eqnarray}
X_j & = & |H\rangle_j \langle V|+|V\rangle_j \langle H|, \\
Y_j & = & i \left(|V\rangle_j \langle H|-|H\rangle_j \langle V|\right), \\
Z_j & = & |H\rangle_j \langle H|-|V\rangle_j \langle V|,
\end{eqnarray}
and three for path degrees of freedom,
\begin{eqnarray}
x_j & = & |u\rangle_j \langle d|+|d\rangle_j \langle u|, \\
y_j & = & i \left(|d\rangle_j \langle u|-|u\rangle_j \langle
d|\right),
\\
z_j & = & |u\rangle_j \langle u|-|d\rangle_j \langle d|.
\end{eqnarray}
Each of these observables can take two values: $-1$ or $1$. Each
observer randomly chooses to measure either a polarization
observable, a path observable, or a polarization observable and a
path observable on his/her photon. The choice of measurement and
the measurement itself on photon-1 are assumed to be spacelike
separated from those on photon-2.

We will prove that these observables satisfy EPR's condition for
LER, namely, {\em ``if, without in any way disturbing a system, we
can predict with certainty (i.e., with probability equal to unity)
the value of a physical quantity, then there exists an element of
physical reality corresponding to this physical
quantity''}~\cite{EPR35}. $Z_1$ and $z_1$ ($Z_2$ and $z_2$) are
EPR's LERs because their values can be predicted with certainty
from spacelike separated measurements of $Z_2$ and $z_2$ ($Z_1$
and $z_1$), respectively, because state~(\ref{benasque}) satisfies
the following equations:
\begin{eqnarray}
Z_1 Z_2 |\psi \rangle & = & |\psi \rangle, \\
z_1 z_2 |\psi \rangle & = & |\psi \rangle.
\end{eqnarray}
$X_1$ and $x_1$ ($X_2$ and $x_2$) are EPR's LERs because their
values can be predicted with certainty from spacelike separated
measurements of $X_2 z_2$ and $Z_2 x_2$ ($X_1 z_1$ and $Z_1 x_1$),
respectively, because state~(\ref{benasque}) satisfies
\begin{eqnarray}
X_1 X_2 z_2 |\psi \rangle & = & |\psi \rangle, \label{equno} \\
x_1 Z_2 x_2 |\psi \rangle & = & |\psi \rangle, \label{eqdos} \\
X_1 z_1 X_2 |\psi \rangle & = & |\psi \rangle, \label{eqtres} \\
Z_1 x_1 x_2 |\psi \rangle & = & |\psi \rangle. \label{eqcuatro}
\end{eqnarray}
Analogously, $Y_1$ and $y_1$ ($Y_2$ and $y_2$) are EPR's LERs
because state~(\ref{benasque}) satisfies
\begin{eqnarray}
Y_1 Y_2 z_2 |\psi \rangle & = & -|\psi \rangle, \label{eqcinco} \\
y_1 Z_2 y_2 |\psi \rangle & = & -|\psi \rangle, \label{eqseis} \\
Y_1 z_1 Y_2 |\psi \rangle & = & -|\psi \rangle, \label{eqsiete} \\
Z_1 y_1 y_2 |\psi \rangle & = & -|\psi \rangle. \label{eqocho}
\end{eqnarray}

We will prove that two compatible observables on the same photon,
like $X_1$ and $z_1$, are independent EPR's LERs in the sense that
the measurement of one of them does not change the value of the
other (and therefore there is no need for any additional
assumptions beyond EPR's condition for LER itself;
see~\cite{Cabello03} for a similar discussion). A suitable
measurement of $X_1$ does not change $v(x_1)$ because $v(x_1)$ can
be predicted with certainty from a spacelike separated measurement
of $Z_2$ and $x_2$, see Eq.~(\ref{eqdos}), and this prediction is
not affected by whether $X_1$ is measured before $x_1$, or $X_1$
and $x_1$ are jointly measured. Therefore, EPR's condition is
enough to guarantee that $x_1$ has a LER [i.e., a value $v(x_1)$]
which does not change with a measurement of $X_1$. A similar
reasoning applies to any other local observable involved in the
proof.

In addition, state~(\ref{benasque}) satisfies the following
equations:
\begin{eqnarray}
X_1 x_1 Y_2 y_2 |\psi \rangle & = & |\psi \rangle, \label{eqnueve} \\
X_1 y_1 Y_2 x_2 |\psi \rangle & = & |\psi \rangle, \label{eqdiez} \\
Y_1 x_1 X_2 y_2 |\psi \rangle & = & |\psi \rangle, \label{eqonce} \\
Y_1 y_1 X_2 x_2 |\psi \rangle & = & |\psi \rangle. \label{eqdoce}
\end{eqnarray}
To be consistent with Eqs.~(\ref{equno})--(\ref{eqdoce}), local
realistic theories predict the following relations between the
values of the LERs:
\begin{eqnarray}
v(X_1) & = & v(X_2) v(z_2),
\label{valuno} \\
v(x_1) & = & v(Z_2) v(x_2),
\label{valdos} \\
v(X_1) v(z_1) & = & v(X_2),
\label{valtres} \\
v(Z_1) v(x_1) & = & v(x_2),
\label{valcuatro} \\
v(Y_1) & = & -v(Y_2) v(z_2),
\label{valcinco} \\
v(y_1) & = & -v(Z_2) v(y_2),
\label{valseis} \\
v(Y_1) v(z_1) & = & -v(Y_2),
\label{valsiete} \\
v(Z_1) v(y_1) & = & -v(y_2),
\label{valocho} \\
v(X_1) v(x_1) & = & v(Y_2) v(y_2),
\label{valnueve} \\
v(X_1) v(y_1) & = & v(Y_2) v(x_2),
\label{valdiez} \\
v(Y_1) v(x_1) & = & v(X_2) v(y_2),
\label{valonce} \\
v(Y_1) v(y_1) & = & v(X_2) v(x_2).
\label{valdoce}
\end{eqnarray}
However, it is impossible to assign the values $-1$ or $1$ to the
observables in a way consistent with
Eqs.~(\ref{valuno})--(\ref{valdoce}), and therefore the
predictions of quantum mechanics cannot be reproduced by EPR's
LERs. Indeed, the assignation is impossible even for each of eight
possible subsets of four equations. For instance, the product of
Eqs.~(\ref{valuno}) and (\ref{valcinco}) [or the product of
Eqs.~(\ref{valtres}) and (\ref{valsiete})] leads to $v(X_1) v(Y_1)
= -v(X_2) v(Y_2)$; while the product of Eqs.~(\ref{valnueve}) and
(\ref{valonce}) [or the product of Eqs.~(\ref{valdiez}) and
(\ref{valdoce})] leads to $v(X_1) v(Y_1) = v(X_2) v(Y_2)$.
Analogously, the product of Eqs.~(\ref{valdos}) and
(\ref{valseis}) [or the product of Eqs.~(\ref{valcuatro}) and
(\ref{valocho})] leads to $v(x_1) v(y_1) = -v(x_2) v(y_2)$; while
the product of Eqs.~(\ref{valnueve}) and (\ref{valdiez}) [or the
product of Eqs.~(\ref{valonce}) and (\ref{valdoce})] leads to
$v(x_1) v(y_1) = v(x_2) v(y_2)$. Note that if we explicitly write
down the eight sets, the four
Eqs.~(\ref{valnueve})--(\ref{valdoce}) would appear twice as
frequently as the eight Eqs.~(\ref{valuno})--(\ref{valocho}).

In a real experiment, measurements are imperfect and the observed
correlation functions fail to attain the values assumed in the
ideal case. Therefore, it is convenient to translate the
contradiction of the AVN proof into a Bell's inequality. This
inequality naturally follows from the observation that the
relevant features of the AVN proof derive from the fact that
state~(\ref{benasque}) is an eigenstate of the operator
\begin{eqnarray}
\beta & = &
X_1 X_2 z_2 + x_1 Z_2 x_2 + X_1 z_1 X_2 + Z_1 x_1 x_2 \nonumber \\
& & -Y_1 Y_2 z_2 - y_1 Z_2 y_2 - Y_1 z_1 Y_2 - Z_1 y_1 y_2 \nonumber \\
& & + 2 X_1 x_1 Y_2 y_2 + 2 X_1 y_1 Y_2 x_2 + 2 Y_1 x_1 X_2 y_2 \nonumber \\
& & + 2 Y_1 y_1 X_2 x_2.
\label{Belloperator}
\end{eqnarray}
As can be easily checked, in any model based on LERs the expected
value of $\beta$ must satisfy
\begin{equation}
|\langle\beta\rangle| \le 8. \label{BellCabello05}
\end{equation}
However, the quantum prediction for the state~(\ref{benasque}) is
\begin{equation}
\left\langle {\psi }\right| \beta \left|{\psi } \right\rangle =
16,
\end{equation}
which is indeed the maximum possible violation of
inequality~(\ref{BellCabello05}). The difference between the
maximal violation of the Bell's inequality and its upper bound is
$16-8 = 8$ for the inequality presented here, while it is just $2
\sqrt{2}-2 \approx 0.8$ for the CHSH inequality~\cite{CHSH69},
$4-2=2$ for the three-qubit version of Mermin's
inequality~\cite{Mermin90c}, and $9-7=2$ for the Bell's inequality
derived from the two-observer AVN proof~\cite{Cabello01a}.

The simplest way to estimate the detection efficiency required to
avoid the detection loophole for a Bell experiment based on this
AVN proof, and a good estimation of the required efficiency for a
test of the inequality (\ref{BellCabello05}), is to see it as a
game in the spirit of Vaidman's game~\cite{Vaidman99} and
Brassard's ``quantum pseudo-telepathy''~\cite{Brassard03}.
Consider a team of two players, Alice and Bob, each of them
isolated in a booth. Each of them is asked one out of eight
possible questions: (i)~``What are $v(X)$ and $v(z)$?,''
(ii)~``What are $v(Z)$ and $v(x)$?,'' (iii)~``What are $v(Y)$ and
$v(z)$?,'' (iv)~``What are $v(Z)$ and $v(y)$?,'' (v)~``What are
$v(X)$ and $V(x)$?,'' (vi)~``What are $v(X)$ and $v(y)$?,''
(vii)~``What are $v(Y)$ and $v(x)$?,'' and (viii)~``What are
$v(Y)$ and $v(y)$?'' If one player is asked a question from (i) to
(iv), then the other is asked the same question; if one is asked
(v), the other is asked (viii); if one is asked (vi), the other is
asked (vii). Therefore, the possible scenarios are (i)-(i),
meaning that both Alice and Bob are asked (i), (ii)-(ii),
(iii)-(iii), (iv)-(iv), (v)-(viii), (vi)-(vii), (vii)-(vi), and
(viii)-(v). Each player must give one of the following answers:
``$-1$ and $-1$,'' ``$-1$ and $1$,'' ``$1$ and $-1$,'' or ``$1$
and $1$.'' Since $v(X)$ represents a LER, Alice's answer to ``What
is $v(X)$?'' must be the same regardless of the scenario in which
is asked. The same applies for all 12 LERs used in the game. Alice
and Bob win if the product of the answers satisfies the
corresponding equation in Eqs.~(\ref{valuno})--(\ref{valdoce}).
Let us assume that all questions are asked with the same
frequency. This is equivalent to assuming that, from the
12~possible scenarios considered in
Eqs.~(\ref{valuno})--(\ref{valdoce}), those of
Eqs.~(\ref{valnueve})--(\ref{valdoce}) occur twice as frequently
than those of Eqs.~(\ref{valuno})--(\ref{valocho}). Assuming this,
it is easy to see that an optimal classical strategy allows the
players to win this game in~$3/4$ of the rounds. For instance, a
simple optimal classical strategy is that the players always use
the following set of local answers:
\begin{eqnarray}
G := \left\{ \begin{array}{cc|cc}
v(X_1) & v(x_1) & v(X_2) & v(x_2) \\
v(Y_1) & v(y_1) & v(Y_2) & v(y_2) \\
v(Z_1) & v(z_1) & v(Z_2) & v(z_2)
\end{array} \right\}
= \left\{ \begin{array}{cc|cc}
1 & 1 & 1 & 1 \\
1 & 1 & 1 & 1 \\
1 & 1 & 1 & 1
\end{array} \right\}.
\end{eqnarray}
This strategy always wins except for scenarios (iii)-(iii) and
(iv)-(iv) [i.e., it satisfies all
Eqs.~(\ref{valuno})--(\ref{valdoce}), except
Eqs.~(\ref{valcinco})--(\ref{valocho})]. However, the players can
win all the rounds if they share pairs of photons in the
state~(\ref{benasque}) and give as answers the results of the
corresponding measurements [i.e., if one is asked question (i),
he/she gives as answers the results of measuring $X$ and $z$ on
his/her photon].

In a real experiment to test the quantum predictions, the low
efficiency of detectors opens the possibility that nondetections
correspond to local instructions like ``if $X$ is measured, then
the photon will not activate the detector.'' This allows us to
construct a model with local instructions which simulates the
observed data by taking advantage of those rounds in which one
photon goes undetected.

Therefore, to estimate the efficiency required for a loophole-free
test consider a modified version of the previous game, including
the possibility of each player not answering in a fraction
$1-\eta$ of the rounds. If any of the players gives no answers,
that round is not taken into account. This new rule opens the
possibility of the players also sharing a fraction of sets of
local instructions like
\begin{eqnarray}
B_1 := \left\{ \begin{array}{cc|cc}
1 & 1 & 1 & 1 \\
0 & 0 & 1 & 1 \\
1 & 1 & 1 & 1
\end{array} \right\},
\end{eqnarray}
or
\begin{eqnarray}
B_2 := \left\{ \begin{array}{cc|cc}
1 & 1 & 1 & 1 \\
1 & 1 & 0 & 0 \\
1 & 1 & 1 & 1
\end{array} \right\},
\end{eqnarray}
where the $0$s in $B_1$ means that, if Alice and Bob are using a
set $B_1$, Alice will not give any answer to questions which
include ``What is $v(Y_1)$?'' or ``What is $v(y_1)$?,'' i.e., to
questions (iii), (iv), (vi), (vii), and (viii). Analogously, the
$0$s in $B_2$ means that, if Alice and Bob are using a set $B_2$,
Bob will not answer questions which include ``What is $v(Y_2)$?''
or ``What is $v(y_2)$?''

Suppose the players are using sets of predefined answers (i.e.,
suppose the observed data are adequately described by a local
realistic theory). For instance, sets like $G$ with a frequency
$1-p$, sets like $B_1$ with a frequency $p/2$, and sets like $B_2$
with a frequency $p/2$. This $p$ is related to the efficiency of
the photodetector corresponding to photon-$j$, $\eta_j$, by the
relation
\begin{equation}
\eta_j = 1 - p + \frac{p}{2} f_j + \frac{p}{2}, \label{eta}
\end{equation}
where $f_1$ ($f_2$) is the probability that Alice (Bob) answers
[i.e., she (he) does not get the instruction $0$ in her (his) set] when
they are using a $B_1$ ($B_2$) set. In our case, $f_j = 3/8$.

Let us calculate the minimum detection efficiency required to
discard the possibility that nature is using this particular set
of predefined answers. To emulate the quantum probability of
winning the game ($P_Q = 1$ in our case), the minimum $p$ must
satisfy
\begin{equation}
P_Q = (1-p) P_G + \frac{p}{2} P_{B_1} + \frac{p}{2} P_{B_2},
\label{PQ}
\end{equation}
where $P_G$ is the probability of winning the game when the
players use a $G$ set, and $P_{B_j}$ is the probability of winning
when the players use a $B_j$ set and both answer the questions. In
our example, $P_G=3/4$ and $P_{B_j}=1$. Introducing the values in
Eqs.~(\ref{eta}) and (\ref{PQ}), we arrive at the conclusion that
our model can simulate the quantum predictions if $\eta_j \le
11/16 \approx 0.69$. An exhaustive examination of all possible
sets like $G$ and $B_j$ shows that the previously presented model
is indeed optimal and, therefore, we conclude that LERs cannot
simulate the quantum predictions if
\begin{equation}
\eta_j > 11/16.
\end{equation}
If we do a similar analysis for a similar game based only on
Eqs.~(\ref{valuno}), (\ref{valcinco}), (\ref{valnueve}), and
(\ref{valonce}), we arrive at the conclusion that closing the
detection hole in this case would require $\eta_j
> 3/4$~\cite{Cabello05a}.




This work was sparked by some experiments made in
Hefei~\cite{YZZYZZCP05} and Rome~\cite{BCDM05} and a talk given by
P.G.~Kwiat in Vienna, and was initiated during the Benasque Center
for Science Quantum Information Workshop. The author thanks
A.~Broadbent, E.~Galv\~{a}o, P.G.~Kwiat, A.~Lamas-Linares,
J.-\AA.~Larsson, C.-Y.~Lu, S.~Lu, E.~Santos, and H.~Weinfurter for
useful comments, and acknowledges support by Projects
Nos.~BFM2002-02815 and~FQM-239.




\begin{thebibliography}{99}


\bibitem{EPR35}
A.~Einstein, B.~Podolsky, and N.~Rosen,
Phys. Rev. {\bf 47}, 777 (1935).

\bibitem{Bell64}
J.S.~Bell,
Physics (Long Island City, NY) {\bf 1}, 195 (1964).

\bibitem{CHSH69}
J.F.~Clauser, M.A.~Horne, A.~Shimony, and R.A.~Holt,
Phys. Rev. Lett. {\bf 23}, 880 (1969).


\bibitem{FC72}
S.J.~Freedman and J.F.~Clauser,
Phys. Rev. Lett. {\bf 28}, 938 (1972);
E.S.~Fry and R.C.~Thompson,
{\em ibid.} {\bf 37}, 465 (1976);
A.~Aspect, J.~Dalibard, and G.~Roger,
{\em ibid.} {\bf 49}, 1804 (1982);
Y.H.~Shih and C.O.~Alley,
{\em ibid.} {\bf 61}, 2921 (1988);
Z.Y.~Ou and L.~Mandel,
{\em ibid.} {\bf 61}, 50 (1988);
Z.Y.~Ou, S.F.~Pereira, H.J.~Kimble, and K.C.~Peng,
{\em ibid.} {\bf 68}, 3663 (1992);
P.R.~Tapster, J.G.~Rarity, and P.C.M.~Owens,
{\em ibid.} {\bf 73}, 1923 (1994);
P.G.~Kwiat, K.~Mattle, H.~Weinfurter, A.~Zeilinger, A.V.~Sergienko,
and Y.H.~Shih,
{\em ibid.} {\bf 75}, 4337 (1995);
W.~Tittel, J.~Brendel, H.~Zbinden, and N.~Gisin,
{\em ibid.} {\bf 81}, 3563 (1998).

\bibitem{WJSHZ98}
G.~Weihs, T.~Jennewein, C.~Simon, H.~Weinfurter,
and A.~Zeilinger,
Phys. Rev. Lett. {\bf 81}, 5039 (1998).

\bibitem{RKVSIMW01}
M.A.~Rowe, D.~Kielpinski, V.~Meyer,
C.A.~Sackett, W.M.~Itano, C.~Monroe, and D.J.~Wineland,
Nature (London) {\bf 409}, 791 (2001).


\bibitem{Bell81}
J.S.~Bell,
J. Phys. C {\bf 2}, 41 (1981);
E.~Santos,
Phys. Lett. A {\bf 200}, 1 (1995).

\bibitem{Pearle70}
P.M. Pearle,
Phys. Rev. D {\bf 2}, 1418 (1970);
E.~Santos,
Phys. Rev. A {\bf 46}, 3646 (1992).


\bibitem{CH74}
J.F.~Clauser and M.A.~Horne,
Phys. Rev. D {\bf 10}, 526 (1974);
N.D.~Mermin,
Ann. N. Y. Acad. Sci. {\bf 480}, 422 (1986);
A.~Garg and N.D.~Mermin,
Phys. Rev. D {\bf 35}, 3831 (1987);
J.-\AA.~Larsson,
Phys. Rev.~A {\bf 57}, 3304 (1998).


\bibitem{LS81}
T.K.~Lo and A.~Shimony,
Phys. Rev. A {\bf 23}, 3003 (1981);
S.F.~Huelga, M.~Ferrero, and E.~Santos,
{\em ibid.} {\bf 51}, 5008 (1995);
E.S.~Fry, T.~Walther, and S.~Li,
{\em ibid.} {\bf 52}, 4381 (1995);
M.~Freyberger, P.K.~Aravind, M.A.~Horne, and A.~Shimony,
{\em ibid.} {\bf 53}, 1232 (1996);
C.~Simon and W.T.M.~Irvine,
Phys. Rev. Lett. {\bf 91}, 110405 (2003);
H.~Nha and H.J.~Carmichael,
Phys. Rev. Lett. {\bf 93}, 020401 (2004);
R.~Garc\'{\i}a-Patr\'{o}n, J.~Fiur\'{a}\v{s}ek, N.J.~Cerf,
J.~Wenger, R.~Tualle-Brouri, and P.~Grangier,
{\em ibid.} {\bf 93}, 130409 (2004).

\bibitem{KESC94}
P.G.~Kwiat, P.H.~Eberhard, A.M.~Steinberg, and R.Y.~Chiao,
Phys. Rev. A {\bf 49}, 3209 (1994).


\bibitem{Kwiat05}
P.G.~Kwiat (private communication).


\bibitem{Eberhard93}
P.H.~Eberhard,
Phys. Rev.~A {\bf 47}, R747 (1993);
A.~Garuccio,
{\em ibid.} {\bf 52}, 2535 (1995).


\bibitem{CG97}
A.~Cabello and G.~Garc\'{\i}a-Alcaine,
J. Phys.~A {\bf 30}, 725 (1997).


\bibitem{GHZ89}
D.M.~Greenberger, M.A.~Horne, and A.~Zeilinger,
in {\em Bell's Theorem, Quantum Theory, and Conceptions of the
Universe}, edited by M.~Kafatos (Kluwer Academic, Dordrecht,
Holland, 1989), p. 69;
D.M.~Greenberger, M.A.~Horne, A.~Shimony, and A.~Zeilinger,
Am. J. Phys. {\bf 58}, 1131 (1990).


\bibitem{Cabello01a}
A. Cabello,
Phys. Rev. Lett. {\bf 86}, 1911 (2001); {\bf 87}, 010403 (2001).


\bibitem{VGG03}
W.~van Dam, P.~Grunwald, and R.D.~Gill,
e-print quant-ph/0307125.


\bibitem{Mermin90a}
N.D.~Mermin,
Phys. Today {\bf 43}(6), 9 (1990);
Am. J. Phys. {\bf 58}, 731 (1990).


\bibitem{Larsson98}
J.-\AA.~Larsson,
Phys. Rev.~A {\bf 57}, R3145 (1998); {\bf 59}, 4801 (1999).


\bibitem{Cabello05a}
A.~Cabello,
Phys. Rev. Lett. {\bf 95}, 210401 (2005).


\bibitem{CPZBZ03}
Z.-B.~Chen, J.-W.~Pan, Y.-D.~Zhang, \v{C}.~Brukner, and A.~Zeilinger,
Phys. Rev. Lett. {\bf 90}, 160408 (2003).

\bibitem{YZZYZZCP05}
T.~Yang, Q.~Zhang, J.~Zhang, J.~Yin, Z.~Zhao, M.~\.{Z}ukowski,
Z.-B.~Chen, and J.-W.~Pan,
e-print quant-ph/0502085.

\bibitem{BCDM05}
M.~Barbieri, C.~Cinelli, F.~De Martini, and P.~Mataloni,
e-print quant-ph/0505098.


\bibitem{Kwiat97}
P.G.~Kwiat,
J. Mod. Opt. {\bf 44}, 2173 (1997);
P.G. Kwiat and H. Weinfurter,
Phys. Rev.~A {\bf 58}, R2623 (1998).


\bibitem{Cabello03}
A.~Cabello,
Phys. Rev. Lett. {\bf 90}, 258902 (2003).


\bibitem{Mermin90c}
N.D.~Mermin,
Phys. Rev. Lett. {\bf 65}, 1838 (1990).


\bibitem{Vaidman99}
L.~Vaidman,
Found. Phys. {\bf 29}, 615 (1999).

\bibitem{Brassard03}
G.~Brassard, A.~Broadbent, and A.~Tapp,
e-print quant-ph/0407221.


\end{thebibliography}
\end{document}